\documentclass[aps,superscriptaddress,twocolumn,dvips]{revtex4}
\usepackage{feynmp}
\usepackage{amsmath}
\usepackage{amssymb}
\usepackage{epsfig}
\usepackage{graphicx}
\usepackage{subfigure}
\usepackage{hyperref}
\usepackage{bbm}
\usepackage{color}
\usepackage{longtable}
\usepackage{booktabs}
\usepackage{multirow}

\newcommand{\Rmnum}[1]{\expandafter\@slowromancap\romannumeral #1@}
\allowdisplaybreaks[3]
\begin{document}

\title{Polarization, plasmon, and Debye screening in doped 3D ani-Weyl semimetal}

\author{Jing-Rong Wang}
\affiliation{Anhui Province Key Laboratory of Condensed Matter
Physics at Extreme Conditions, High Magnetic Field Laboratory of the
Chinese Academy of Science,Hefei 230031, Anhui, China}
\author{Guo-Zhu Liu}
\altaffiliation{Corresponding author: gzliu@ustc.edu.cn}
\affiliation{Department of Modern Physics, University of Science and
Technology of China, Hefei, Anhui 230026, P. R. China}
\author{Chang-Jin Zhang}
\altaffiliation{Corresponding author: zhangcj@hmfl.ac.cn}
\affiliation{Anhui Province Key Laboratory of Condensed Matter
Physics at Extreme Conditions, High Magnetic Field Laboratory of the
Chinese Academy of Science,Hefei 230031, Anhui,
China}\affiliation{Collaborative Innovation Center of Advanced
Microstructures, Nanjing University, Nanjing 210093, P. R. China}

\begin{abstract}
We compute the polarization function in a doped three-dimensional
anisotropic-Weyl semimetal, in which the fermion energy dispersion
is linear in two components of the momenta and quadratic in the
third. Through detailed calculations, we find that the long
wavelength plasmon mode depends on the fermion density $n_e$ in the
form $\Omega_{p}^{\bot}\propto n_{e}^{3/10}$ within the basal plane
and behaves as $\Omega_{p}^{z}\propto n_{e}^{1/2}$ along the third
direction. This unique characteristic of the plasmon mode can be
probed by various experimental techniques, such as electron
energy-loss spectroscopy. The Debye screening at finite chemical
potential and finite temperature is also analyzed based on the
polarization function.
\end{abstract}

\maketitle


\section{Introduction}

Studying the intriguing properties of various semimetals have been
one of the core subjects in condensed matter physics for more than
one decade \cite{Vafek14, Wehling14, Wan11, Burkov16, Yan17,
Hasan17, Burkov17, Armitage17, Weng16, FangChen16}, starting from
the successful fabrication of monolayer graphene \cite{CastroNeto09,
Kotov12}. The surface state of three-dimensional (3D) topological
insulator bears strong similarity to graphene, and both of these two
systems are classified as two-dimensional (2D) Dirac semimetal (DSM)
\cite{Hasan10, Qi11}. It is currently clear that there are a variety
of semimetals materials, including 3D DSM \cite{Vafek14, Wehling14},
3D Weyl semimetal (WSM) \cite{Wan11, Burkov16, Yan17,Hasan17,
Burkov17, Armitage17}, and nodal line semimetal (NLSM) \cite{Weng16,
FangChen16}. It is interesting that some of such semimetals provide
a nice platform to realize the interesting concepts of chiral
anomaly by probing the negative magnetoresistance
\cite{HuangChenGFGroup15, ZhangChengLong16}. In addition, there exit
2D semi-DSM \cite{Hasegawa06, Dietl08, Montambaux09B, Pardo09,
Banerjee09, Banerjee12, Dora13, Pyatkovskiy16, Isobe16, Cho16,
WangLiuZhang17}, 3D double-WSM \cite{XuFang11, Fang12, Yang14A,
Huang16, Lai15, Jian15, WangLiuZhang16, BeraRoy16, Sbierski17,
Roy17, Ahn16, Ahn17, Park17}, 3D triple-WSM \cite{Yang14A,
WangLiuZhang16, Roy17, Ahn16, Ahn17, Park17, LiuZunger17, Zhang16,
WangLiuZhang17B}, and 3D anisotropic-WSM  (ani-WSM) \cite{Yang14A,
Yang13, Yang14B, Moon14}.

In this paper, we pay attention to 3D ani-WSM where the Weyl fermion
spectrum displays linear dependence on two components of the momenta
and quadratic dependence on the third \cite{Yang14A, Yang13,
Yang14B, Moon14}, namely
\begin{eqnarray}
E = \pm\sqrt{v^{2}\left(k_{x}^{2} + k_{y}^{2}\right) +
A^{2}k_{z}^{4}}. \label{Eq:AWDispersion}
\end{eqnarray}
Such 3D ani-WSM might be produced at the quantum critical point
(QCP) between a normal insulator and a WSM, or at the QCP between a
normal insulator and a topological insulator in 3D
noncentrosymmetric systems \cite{Yang13}. For 3D ani-WSM, the
chirality of the band-touching point is zero \cite{Yang13}. First
principle calculations suggested that the topological quantum phase
transition (QPT) from normal insulator to topological insulator in a
noncentrosymmetric system can be tuned by applying pressure to BiTeI
\cite{Bahramy12}. This theoretical prediction was subsequently
confirmed by experiments carried out by means of $x$-ray powder
diffraction and infrared spectroscopy techniques \cite{Xi13}.

It is also proposed \cite{Yang14A} that a semimetallic state in
which the fermion excitations have the dispersion of
Eq.~(\ref{Eq:AWDispersion}) can emerge at the QCP between normal
insulator and topological 3D DSM, or the QCP between 3D DSM and weak
topological insulator or topological crystalline insulator. The
analysis made by Yuan \emph{et al.} \cite{Yuan17} showed that the
fermion dispersion given by Eq.~(\ref{Eq:AWDispersion}) may be
realized in ZrTe$_{5}$ at the QCP between insulator phase and DSM
phase. Remarkably, recent quantum oscillation measurements provided
important clue for the existence of such type of fermionic
excitations in ZrTe$_{5}$ under pressure \cite{ZhangJingLei17}.

In an intrinsic semimetal, the chemical potential is exactly zero,
and the density of states (DOS) vanishes at the Fermi level. As a
result, the dynamically screened Coulomb interaction between the
fermion excitations is still long-ranged. Extensive renormalization
group (RG) \cite{Shankar94} analysis revealed that the long-range
Coulomb interaction in various intrinsic semimetals might be
marginally irrelevant \cite{Kotov12, Isobe16, Cho16, WangLiuZhang17,
Goswami11, Lai15, Jian15, Zhang16, WangLiuZhang16}, relevant
\cite{Herbut14, JanssenSeries}, or irrelevant \cite{Yang14B, Huh16},
determined by the specific fermion dispersion and the spatial
dimension of the system. In the special case of 3D ani-WSM, the
Coulomb interaction is found to be irrelevant \cite{Yang14B} and
thus does not significantly modify the low-energy behaviors of free
fermions, which is consistent with the previous work of Abrikosov
\cite{Abrikosov72}.

In realistic semimetal materials, the chemical potential usually
takes a finite value, and its value can be adjusted by changing the
gate voltage. For semimetals defined at a finite chemical potential,
there are undamped collective modes, namely plasmon. The properties
of plasmon is directly related to the energy dispersion of fermion
excitations, and can be experimentally investigated by various
techniques, such as the electron energy-loss spectroscopy
\cite{GiulianiBook, MaierBook}. In recent years, there appeared
certain amount of theoretic studies for the plasmon mode in several
sorts of doped semimetals, including 2D DSM \cite{Wunsch06, Hwang07,
Pyatkovskiy09}, 2D semi-DSM \cite{Banerjee12, Pyatkovskiy16}, 3D
DSM/WSM \cite{Sarma09, Lv13, Panfilov14, Zhou15}, 3D NLSM
\cite{YanZhongbo16, Rhim16}, 3D multi-WSMs \cite{Ahn16}, and some
other novel fermion systems \cite{Stauber14, Chang14}. However, a
systematic study for the plasmon in the case of doped 3D ani-WSM is
still lacking.

In this paper, we calculate the polarization function in the context
of doped 3D ani-WSM, and then analyze the asymptotic behavior of
plasmon in the long wavelength limit. It is found that the plasmon
within the basal plane exhibits the behavior
$\Omega_{p}^{\bot}\propto n_{e}^{\frac{3}{10}}$, and the one along
the third direction is given by $\Omega_{p}^{z}\propto
n_{e}^{\frac{1}{2}}$, where $n_{e}$ is the density of fermions.
These behaviors could be probed by experiments, which is expected to
offer important evidence for the existence of fermions with the
dispersion given by Eq.~(\ref{Eq:AWDispersion}). On the basis of the
polarization, the Debye screening at finite chemical potential and
finite temperature is also analyzed.

The rest of the paper is organized as follows. The model considered
by us is presented in Sec.~\ref{Sec:Model}. In
Sec.~\ref{Sec:GeneralPola}, the polarization function is calculated
and presented in the form a general expression. The plasmon in doped
3D ani-WSM is analyzed in Sec.~\ref{Sec:Plasmon}. In
Sec.~\ref{Sec:DebyeScreening}, we analyze the behaviors of the Debye
screening at finite chemical potential and finite temperature. We
summarize our results in Sec.~\ref{Sec:Summary}, and present in the
Appendices the calculational details for the derivation of the
polarization function in the long wavelength limit, the integration over
the azimuthal angle, and the analysis of Debye screening.

\section{Model and Feynman rules \label{Sec:Model}}

For a doped 3D ani-WSM, the free fermion action is given by
\begin{eqnarray}
S_{f} = \int d\tau d^3\mathbf{x} \psi^{\dag}\left(\partial_{\tau} +
\mu + \mathcal{H}\right)\psi,
\end{eqnarray}
where $\psi$ represents the spinor field and $\mu$ is the chemical
potential. The Hamiltonian density takes the from
\begin{eqnarray}
\mathcal{H} = iv\partial_{x}\sigma_{x} + iv\partial_{y}\sigma_{y} +
A\partial_{z}^{2}\sigma_{z}
\end{eqnarray}
with $\sigma_{x,y,z}$ being the Pauli matrices. Here, $v$ and $A$
are two model parameters. The Coulomb interaction is described by
\begin{eqnarray}
S_{\mathrm{C}} = \int d\tau d^3\mathbf{x}d^3\mathbf{x}'
\left(\psi^{\dag}\psi\right)_{\tau,\mathbf{x}}
\frac{e^2}{2\kappa\left|\mathbf{x}-\mathbf{x}'\right|}
\left(\psi^{\dag}\psi\right)_{\tau,\mathbf{x'}},
\end{eqnarray}
where $e$ is the electric charge and $\kappa$ the dielectric
constant.

The fermion propagator is written in the Matsubara formalism:
\begin{eqnarray}
G_{0}(i\omega_{n},\mathbf{k}) = \frac{1}{i\omega_{n} + \mu -
\left(vk_{x}\sigma_{x}+vk_{y}\sigma_{y} +
Ak_{z}^{2}\sigma_{z}\right)}, \label{Eq:PropagatorMatsubara}
\end{eqnarray}
where $\omega_{n} = (2n+1)\pi T$. Making analytical continuation
$i\omega_{n}\rightarrow \omega + i\eta$ leads to the retarded
fermion propagator:
\begin{eqnarray}
G_{0}^{\mathrm{ret}}(\omega,\mathbf{k}) = \frac{1}{\omega + \mu -
\left(vk_{x}\sigma_{x}+vk_{y}\sigma_{y} +
Ak_{z}^{2}\sigma_{z}\right)+i\eta}. \label{Eq:PropagatorRetarded}
\end{eqnarray}
The bare Coulomb interaction can be written as
\begin{eqnarray}
V_{0}\left(\mathbf{q}\right) = \frac{4\pi e^{2}}{\kappa
|\mathbf{q}|^{2}}. \label{Eq:BareCoulomb}
\end{eqnarray}
After including the dynamical screening caused by the collective
particle-hole excitations, we write down the dressed retarded
Coulomb interaction
\begin{eqnarray}
V^{\mathrm{ret}}\left(\Omega,\mathbf{q}\right) =
\frac{V_{0}(\mathbf{q})}{\epsilon_{r}(\Omega,\mathbf{q})},
\end{eqnarray}
which is obtained by using the random-phase approximation (RPA). The
dielectric function $\epsilon_{r}(\Omega,\mathbf{q})$ is defined as
\begin{eqnarray}
\epsilon_{r}(\Omega,\mathbf{q}) = 1 +
V_{0}(\mathbf{q})\Pi^{\mathrm{ret}}(\Omega,\mathbf{q}),
\end{eqnarray}
where $\Pi^{\mathrm{ret}}(\Omega,\mathbf{q})$ is the retarded
polarization function. The collective plasmon mode is determined by
the condition
\begin{eqnarray}
\epsilon_{r}(\Omega_{p},\mathbf{q}) = 1 + V_{0}(\mathbf{q})
\Pi^{\mathrm{ret}}(\Omega_{p},\mathbf{q}) = 0.
\label{Eq:PlasmonDefinition}
\end{eqnarray}
In the following, we will calculate the polarization function and
then analyze the asymptotic behavior of the plasmon mode in the
long wavelength limit.

\section{General expression of polarization function \label{Sec:GeneralPola}}

To the leading order, the polarization function reads
\begin{eqnarray}
\Pi(i\Omega_{m},\mathbf{q}) &=& -N\frac{1}{\beta} \sum_{i\omega_{n}}
\int\frac{d^3\mathbf{k}}{(2\pi)^{3}}
\mathrm{Tr}\left[G_{0}(i\omega_{n},\mathbf{k})\right.\nonumber
\\
&&\left.\times G_{0}\left(i\left(\omega_{n} +
\Omega_{m}\right),\mathbf{k}+\mathbf{q}\right)\right],
\label{Eq:PolaExpMainTextA}
\end{eqnarray}
where $\Omega_{m} = 2m\pi T$ and $\beta = \frac{1}{T}$. Utilizing
the standard spectral representation
\begin{equation}
G_0\left(i\omega_n,\mathbf{k}\right) = -\int_{-\infty}^{+\infty}
\frac{d\omega_1}{\pi}\frac{\mathrm{Im}\left[G_0^{\mathrm{ret}}\left(
\omega_1,\mathbf{k}\right)\right]}{i\omega_n-\omega_1},
\end{equation}
we obtain
\begin{eqnarray}
\Pi(i\Omega_{m},\mathbf{q})
&=&-N\int\frac{d^3\mathbf{k}}{(2\pi)^{3}}
\mathrm{Tr}\left[\int_{-\infty}^{+\infty}
\frac{d\omega_1}{\pi}\mathrm{Im}\left[G_0^{\mathrm{ret}}
\left(\omega_1,\mathbf{k}\right)\right]\right.\nonumber
\\
&&\times\left. \int_{-\infty}^{+\infty} \frac{d\omega_2}{\pi}
\mathrm{Im}\left[G_0^{\mathrm{ret}}\left(\omega_2,\mathbf{k} +
\mathbf{q}\right)\right]\right]\nonumber
\\
&&\times\frac{1}{\beta}\sum_{i\omega_{n}}\frac{1}{{i\omega_n -
\omega_1}}\frac{1}{i\omega_n+i\Omega_{m}-\omega_2}.
\label{Eq:PolaExpMainTextB}
\end{eqnarray}
Summing up all frequencies leads to
\begin{equation}
\frac{1}{\beta}\sum_{i\omega_n}\frac{1}{i\omega_n-\omega_1}
\frac{1}{i\omega_n+i\Omega_{m}-\omega_2} =
\frac{n_F\left(\omega_1\right) -
n_F\left(\omega_2\right)}{\omega_1-\omega_2+i\Omega_{m}},
\end{equation}
which then yields
\begin{eqnarray}
\Pi(i\Omega_{m},\mathbf{q}) &=& -N\int
\frac{d^3\mathbf{k}}{(2\pi)^{3}}\mathrm{Tr}\left[\int_{-\infty}^{+\infty}
\frac{d\omega_1}{\pi}\mathrm{Im}\left[G_0^{\mathrm{ret}}
\left(\omega_1,\mathbf{k}\right)\right]\right.\nonumber
\\
&&\left.\times\int_{-\infty}^{+\infty}
\frac{d\omega_2}{\pi}\mathrm{Im}\left[G_0^{\mathrm{ret}}
\left(\omega_2,\mathbf{k} + \mathbf{q}\right)\right]\right]\nonumber
\\
&&\times\frac{n_F\left(\omega_1\right)-n_F\left(\omega_2\right)}{\omega_1
- \omega_2+i\Omega_{m}}. \label{Eq:PolaExpMainTextC}
\end{eqnarray}
The imaginary part of the retarded fermion propagator is then given
by
\begin{eqnarray}
&&\mathrm{Im}\left[G_{0}^{\mathrm{ret}}(\omega,\mathbf{k})\right]\nonumber
\\
&=&-\pi\mathrm{sgn}(\omega+\mu)\left(\omega+\mu+vk_{x}\sigma_{x}+vk_{y}\sigma_{y}
+Ak_{z}^{2}\sigma_{z}\right)\nonumber
\\
&&\times\frac{1}{2E_{\mathbf{k}}}\left[\delta\left(\omega+\mu+E_{\mathbf{k}}\right)
+\delta\left(\omega+\mu-E_{\mathbf{k}}\right)\right],
\label{Eq:ImPartPropagator}
\end{eqnarray}
where the energy
\begin{eqnarray}
E_{\mathbf{k}} = \sqrt{v^{2}k_{\bot}^{2}+A^{2}k_{z}^{4}}.
\end{eqnarray}
Substituting Eq.~(\ref{Eq:ImPartPropagator}) into
Eq.~(\ref{Eq:PolaExpMainTextC}), we find that
\begin{eqnarray}
\Pi(i\Omega_{m},\mathbf{q}) &=& -\frac{N}{16\pi^{3}}
\sum_{\alpha,\alpha'=\pm1}\int d^3\mathbf{k} \left[1 + \alpha\alpha'
\frac{F_{\mathbf{k},\mathbf{q}}}{E_{\mathbf{k}}
E_{\mathbf{k}+\mathbf{q}}}\right]\nonumber
\\
&&\times\frac{n_F\left(\alpha E_{\mathbf{k}}-\mu\right) -
n_F\left(\alpha'E_{\mathbf{k}+\mathbf{q}}-\mu\right)}{\alpha
E_{\mathbf{k}}-\alpha'E_{\mathbf{k}+\mathbf{q}}+i\Omega_{m}},
\end{eqnarray}
where
\begin{eqnarray}
F_{\mathbf{k},\mathbf{q}}&=&v^{2}k_{x}\left(k_{x}+q_{x}\right)
+ v^{2}k_{y}\left(k_{y}+q_{y}\right)\nonumber \\
&&+A^2k_{z}^{2}\left(k_{z}+q_{z}\right)^{2}.
\end{eqnarray}
Here, $n_{F}(E) = \frac{1}{e^{E/T}+1}$ is the Fermi-Dirac
distribution function. We then perform the following analytic
continuation:
\begin{equation}
\frac{1}{x+i\Omega_{m}}\rightarrow \frac{1}{x + \Omega + i\eta}
=P\frac{1}{x+\Omega}-i\pi\delta\left(x+\Omega\right).
\end{equation}
Now the imaginary and real parts of the retarded polarization
function become
\begin{eqnarray}
\mathrm{Im}\left[\Pi^{\mathrm{ret}}(\Omega,\mathbf{q})\right] &=&
\frac{N}{16\pi^{2}}\sum_{\alpha,\alpha'=\pm1}\int
d^3\mathbf{k}\nonumber \\
&&\times\left[1+\alpha\alpha'
\frac{F_{\mathbf{k},\mathbf{q}}}{E_{\mathbf{k}}E_{\mathbf{k} +
\mathbf{q}}}\right]\nonumber \\
&&\times\left[n_F\left(\alpha E_{\mathbf{k}}-\mu\right) -
n_F\left(\alpha'E_{\mathbf{k}+\mathbf{q}} -
\mu\right)\right]\nonumber \\
&&\times\delta\left(\alpha E_{\mathbf{k}} -
\alpha'E_{\mathbf{k}+\mathbf{q}}+\Omega\right),
\label{Eq:ImPolaMainText}
\end{eqnarray}
and
\begin{eqnarray}
\mathrm{Re}\left[\Pi^{\mathrm{ret}}(\Omega,\mathbf{q})\right] &=&
-\frac{N}{16\pi^{3}}\sum_{\alpha,\alpha'=\pm1}P\int
d^3\mathbf{k}\nonumber \\
&&\times\left[1+\alpha\alpha'
\frac{F_{\mathbf{k},\mathbf{q}}}{E_{\mathbf{k}}E_{\mathbf{k} +
\mathbf{q}}}\right]\nonumber
\\
&&\times\left[n_F\left(\alpha E_{\mathbf{k}}-\mu\right) -
n_F\left(\alpha'E_{\mathbf{k}+\mathbf{q}}-\mu\right)\right]\nonumber
\\
&&\times\frac{1}{\alpha E_{\mathbf{k}} -
\alpha'E_{\mathbf{k}+\mathbf{q}}+\Omega}.\label{Eq:RePolaMainText}
\end{eqnarray}
The derivation for the polarization function in the long wavelength
limit is shown in Appendix~\ref{App:PolaAppro}, and the calculation
details for the imaginary and real parts of the retarded
polarization function are presented in
Appendix~\ref{App:PolaAnagleIntegral}. Since the polarization
function is invariant under the transformation $\mu\rightarrow
-\mu$, reflecting the particle-hole symmetry, we will choose $\mu
> 0$ in the subsequent calculations.

\section{Plasmon mode \label{Sec:Plasmon}}

In this section, we analyze the long wavelength plasmon in doped 3D
ani-WSM at zero temperature.

\begin{table*}[htbp]
\caption{Plasmon mode in various doped semimetals. Here, $n_{e}$ is
the fermion density, and the function $n_{e}(\mu)$ describes the
relation between $n_{e}$ and chemical potential $\mu$. The plasmon
mode is characterized by $\Omega_{p}$. \label{Table:PlasmonSummary}}
\begin{center}
\begin{ruledtabular}
\begin{tabular}{lcccc}
\toprule    Material         & Fermion dispersion & $n_{e}(\mu)$ &
$\Omega_{p}$ & Reference
\\ \hline
\midrule 2D Meta         & $E=\pm A|\mathbf{k}|^{2}$
&$n_{e}\propto\mu$ & $\Omega_{p}\propto
n_{e}^{\frac{1}{2}}|\mathbf{q}|^{\frac{1}{2}}$  &
\cite{GiulianiBook, MaierBook, Sarma09}
\\
doped 2D DSM          & $E=\pm v|\mathbf{k}|$  & $n_{e}\propto\mu^{2}$ &
$\Omega_{p}\propto n_{e}^{\frac{1}{4}}|\mathbf{q}|^{\frac{1}{2}}$
&  \cite{Wunsch06, Hwang07}
\\
doped 2D Semi-DSM        &
$E=\pm\sqrt{A^{2}k_{x}^{4}+v^{2}k_{y}^{2}}$ &
$n_{e}\propto\mu^{\frac{3}{2}}$  & $\Omega_{p}^{x}\propto
n_{e}^{\frac{2}{3}}|q_{x}|^{\frac{1}{2}}, \Omega_{p}^{y}\propto
n_{e}^{0}|q_{y}|^{\frac{1}{2}}$ & \cite{Pyatkovskiy16}
\\
3D Meta         & $E=\pm A|\mathbf{k}|^{2}$ & $n_{e}\propto
\mu^{\frac{3}{2}}$ & $\Omega_{p}\propto n_{e}^{\frac{1}{2}}$  &
\cite{GiulianiBook, MaierBook, Sarma09}
\\
doped 3D DSM/WSM          & $E=\pm v|\mathbf{k}|$ &
$n_{e}\propto\mu^{3}$ & $\Omega_{p}\propto n_{e}^{\frac{1}{3}}$  &
\cite{Sarma09, Lv13, Panfilov14, Zhou15}
\\
doped 3D NLSM          &
$E=\pm\sqrt{A^2(k_{\bot}^{2}-k_{F}^{2})^{2}+v^2k_{z}^{2}}$ &
$n_{e}\propto\mu^{2}$& $\Omega_{p}\propto n_{e}^{\frac{1}{4}}$ &
\cite{YanZhongbo16, Rhim16}
\\
doped 3D double-WSM         &
$E=\pm\sqrt{A^{2}k_{\bot}^{4}+v^2k_{z}^{2}}$ & $n_{e}\propto
\mu^{2}$ &$\Omega_{p}^{\bot}\propto n_{e}^{\frac{1}{2}},
\Omega_{p}^{z}\propto n_{e}^{\frac{1}{4}}$ & \cite{Ahn16}
\\
doped 3D triple-WSM         &
$E=\pm\sqrt{B^{2}k_{\bot}^{6}+v^2k_{z}^{2}}$ & $n_{e}\propto
\mu^{\frac{5}{3}}$ &$\Omega_{p}^{\bot}\propto n_{e}^{\frac{3}{5}},
\Omega_{p}^{z}\propto n_{e}^{\frac{1}{5}}$ & \cite{Ahn16}
\\
doped 3D ani-WSM          &
$E=\pm\sqrt{v^2k_{\bot}^{2}+A^2k_{z}^{4}}$ & $n_{e}\propto
\mu^{\frac{5}{2}}$ & $\Omega_{p}^{\bot}\propto n_{e}^{\frac{3}{10}},
\Omega_{p}^{z}\propto n_{e}^{\frac{1}{2}}$ &
\\
\bottomrule
\end{tabular}
\end{ruledtabular}
\end{center}
\end{table*}

As $T \rightarrow 0$, the function $n_{F}(E)$ becomes the step
function $\theta(E)$. In the long wavelength regime with
$\max(vq_{\bot},Aq_{z}^{2})\ll \Omega\ll\mu$, it is easy to find
that $\mathrm{Im}\Pi^{\mathrm{ret}}(\Omega,\mathbf{q}) = 0$, which
implies the existence of undamped plasmon. According to the values
of $\alpha$ and $\alpha'$, we can see that the real part of the
polarization function is divided into four parts, namely
$\mathrm{Re}\Pi_{++}^{\mathrm{ret}}(\Omega,\mathbf{q})$,
$\mathrm{Re}\Pi_{+-}^{\mathrm{ret}}(\Omega,\mathbf{q})$,
$\mathrm{Re}\Pi_{-+}^{\mathrm{ret}}(\Omega,\mathbf{q})$, and
$\mathrm{Re}\Pi_{--}^{\mathrm{ret}}(\Omega,\mathbf{q})$. In the
regime $\max(vq_{\bot},Aq_{z}^{2})\ll \Omega\ll\mu$, it is easy to
verify that $\mathrm{Re}\Pi_{--}^{\mathrm{ret}}(\Omega,\mathbf{q}) =
0$ due to the relation
\begin{eqnarray}
\left[\theta\left(-E_{\mathbf{k}}-\mu\right) -
\theta\left(-E_{\mathbf{k}+\mathbf{q}} - \mu\right)\right]=0.
\end{eqnarray}
As shown in Appendix~\ref{App:PolaAppro}, in
the regime $\max(vq_{\bot},Aq_{z}^{2})\ll \Omega\ll\mu$, we have
\begin{eqnarray}
\mathrm{Re}\Pi_{++}^{\mathrm{ret}}(\Omega,\mathbf{q})
&\approx&-\left(C_{++}^{\bot}\frac{q_{\bot}^{2}\mu^{\frac{3}{2}}}{\Omega^{2}}+
C_{++}^{z}\frac{q_{z}^{2}\mu^{\frac{5}{2}}}{\Omega^{2}}\right).
\\
\mathrm{Re}\Pi_{+-}^{\mathrm{ret}}(\Omega,\mathbf{q})
&\approx&C_{+-}^{\bot}q_{\bot}^{2}+C_{+-}^{z}q_{z}^{2}, \label{Eq:PolaPNMainText}
\\
\mathrm{Re}\Pi_{-+}^{\mathrm{ret}}(\Omega,\mathbf{q})
&\approx&C_{-+}^{\bot}q_{\bot}^{2}+C_{-+}^{z}q_{z}^{2}, \label{Eq:PolaNPMainText}
\end{eqnarray}
where
\begin{eqnarray}
C_{++}^{\bot}&=&\frac{N}{5\pi^{2}\sqrt{A}},
\\
C_{++}^{z}&=&\frac{3N\Gamma\left(\frac{5}{4}\right)
\sqrt{A}}{2\pi^{\frac{3}{2}}\Gamma\left(\frac{11}{4}\right)v^{2}},
\end{eqnarray}
and
\begin{eqnarray}
C_{+-}^{\bot}&=&C_{-+}^{\bot}=\frac{N}{20\pi^2\sqrt{A}}
\left(\frac{1}{\Lambda^{\frac{1}{2}}}-\frac{1}{\mu^{\frac{1}{2}}}\right),
\label{Eq:CNPBotMainText}
\\
C_{+-}^{z}&=&C_{-+}^{z}=\left(\frac{6}{7}-\frac{3\sqrt{\pi}
\Gamma\left(\frac{5}{4}\right)}{2\Gamma\left(\frac{11}{4}\right)}\right)
\frac{N\sqrt{A}}{32v^{2}}\nonumber
\\
&&\times\left(\Lambda^{\frac{1}{2}}-\mu^{\frac{1}{2}}\right).
\label{Eq:CNPZMainText}
\end{eqnarray}
Noticing that
\begin{eqnarray}
\left|\mathrm{Re}\Pi_{++}^{\mathrm{ret}}(\Omega,\mathbf{q})\right|
\gg
\left|\mathrm{Re}\Pi_{+-}^{\mathrm{ret}}(\Omega,\mathbf{q})\right|,
\\
\left|\mathrm{Re}\Pi_{++}^{\mathrm{ret}}(\Omega,\mathbf{q})\right|
\gg
\left|\mathrm{Re}\Pi_{-+}^{\mathrm{ret}}(\Omega,\mathbf{q})\right|,
\end{eqnarray}
which is valid in the regime $\max(vq_{\bot},Aq_{z}^{2})\ll
\Omega\ll\mu$, we finally get
\begin{eqnarray}
\mathrm{Re}\Pi^{\mathrm{ret}}(\Omega,\mathbf{q}) &\approx&
\mathrm{Re}\Pi_{++}^{\mathrm{ret}}(\Omega,\mathbf{q})\nonumber
\\
&\approx&-\left(C_{++}^{\bot}\frac{q_{\bot}^{2}
\mu^{\frac{3}{2}}}{\Omega^{2}}+ C_{++}^{z}
\frac{q_{z}^{2}\mu^{\frac{5}{2}}}{\Omega^{2}}\right).
\label{Eq:PolaApproxMainText}
\end{eqnarray}

As mentioned in Sec.~\ref{Sec:Model}, the plasmon mode  is
determined by Eq.~(\ref{Eq:PlasmonDefinition}), which gives rise to
\begin{eqnarray}
1 - \frac{4\pi e^2\kappa}{q_{\bot}^{2} + q_{z}^{2}}
\left(C_{++}^{\bot}\frac{q_{\bot}^{2}\mu^{\frac{3}{2}}}{\Omega^{2}}
+ C_{++}^{z}\frac{q_{z}^{2}\mu^{\frac{5}{2}}}{\Omega^{2}}\right)=0.
\end{eqnarray}
From this equation, we can get the plasmon mode
\begin{eqnarray}
\Omega_{p} = C_{0}\sqrt{C_{++}^{\bot} \mu^{\frac{3}{2}}\sin^2(\phi)
+ C_{++}^{z}\mu^{\frac{5}{2}}\cos^2(\phi)},
\end{eqnarray}
where $C_{0}=\sqrt{4\pi e^2\kappa}$ and $\phi$ is the angle between
$\mathbf{q}$ and $z$-axis. The plasmon mode within the basal
plane and the one along the third direction can be respectively
written as
\begin{eqnarray}
\Omega_{p}^{\bot} &=& \sqrt{4\pi e^2\kappa C_{++}^{\bot}}
\mu^{\frac{3}{4}}\propto \mu^{\frac{3}{4}},
\\
\Omega_{p}^{z} &=& \sqrt{4\pi e^2\kappa C_{++}^{z}}
\mu^{\frac{5}{4}}\propto \mu^{\frac{5}{4}}.
\end{eqnarray}
According to calculations presented in
Appendix~\ref{App:DensityFermion}, the relation between carrier
density $n_{e}$ and chemical potential $\mu$ is
\begin{eqnarray}
n_{e} = \frac{1}{5\pi^{2}v^{2}\sqrt{A}}\mu^{\frac{5}{2}}.
\end{eqnarray}
It is now straightforward to obtain
\begin{eqnarray}
\Omega_{p}^{\bot} &\propto& n_{e}^{\frac{3}{10}},
\\
\Omega_{p}^{z}&\propto& n_{e}^{\frac{1}{2}},
\end{eqnarray}
which is a unique characteristic of doped 3D ani-WSM, and can be
probed experimentally.

We now would like to compare the plasmon mode in doped 3D ani-WSM
with other analogous semimetals. The behaviors of plasmon mode
obtained in the context of various doped semimetals as well as
traditional metals are summarized in Table
\ref{Table:PlasmonSummary}. It is clear that the plasmon in 3D
fermion systems is always gapped, but the one in 2D fermion systems
is gapless. The difference arises form the fact that the bare
Coulomb interaction has distinct momentum dependence in 2D and 3D.
The bare Coulomb interaction is given by Eq.~(\ref{Eq:BareCoulomb})
in 3D, and has the form
\begin{eqnarray}
V_{0}^{\mathrm{2D}}(\mathbf{q})=\frac{2\pi e^2}{\kappa |\mathbf{q}|}
\end{eqnarray}
in 2D. For a fixed spatial dimension, the behavior of the plasmon
mode is closely related to the energy dispersion of the fermionic
excitations. More concretely, in the isotropic case, the plasmon is
also isotropic and its power takes different value in different
semimetals. In semimetals with anisotropic fermion dispersion, the
plasmon mode is also anisotropic.

\section{Debye screening\label{Sec:DebyeScreening}}

Based on the results obtained in Appendix \ref{App:Screening}, we
find that
\begin{eqnarray}
\mathrm{Re}\Pi^{\mathrm{ret}}(0,0)= \frac{N
T^{\frac{3}{2}}}{2\pi^{2}v^{2}\sqrt{A}}\sum_{\alpha=\pm 1}
\int_{0}^{+\infty}dx\frac{\sqrt{x}}{e^{x+\frac{\alpha\mu}{T}}+1}
\nonumber \\
\end{eqnarray}
in the limit of $\Omega = 0$ and $|\mathbf{q}| = 0$, which
represents the Debye screening induced by finite chemical potential
and finite temperature. At $T > 0$ and $\mu=0$, we have
\begin{eqnarray}
\mathrm{Re}\Pi^{\mathrm{ret}}(0,0) = \frac{1}{4}(2-\sqrt{2})
\zeta\left(\frac{3}{2}\right)\frac{NT^{\frac{3}{2}}}{\pi^{\frac{3}{2}}
v^{2}\sqrt{A}},
\end{eqnarray}
whereas at $T=0$ and $\mu > 0$ we find that
\begin{eqnarray}
\mathrm{Re}\Pi^{\mathrm{ret}}(0,0) = \frac{N}{3\pi^{2}
v^{2}\sqrt{A}}\mu^{\frac{3}{2}}.
\end{eqnarray}
As a consequence of the Debye screening, the dressed Coulomb
interaction becomes short-ranged.

\section{Summary\label{Sec:Summary}}

In summary, we calculate the polarization function in the context of
a doped 3D ani-WSM, and analyze the behavior of the long wavelength
plasmon. We find that the plasmon within the basal plane depends on
the fermion density in the form
$\Omega_{p}^{\bot}=n_{e}^{\frac{3}{10}}$, and that the plasmon along
the third direction behaves as $\Omega_{p}^{z}\propto
n_{e}^{\frac{1}{2}}$. These behaviors can be experimentally detected
and would provide a useful methods to verify the existence of 3D
ani-Weyl fermions. We expect that such plasmon mode could be
confirmed in BiTeI, ZrTe$_{5}$, and other semimetals that host
fermion excitations with dispersion Eq.(1).

\section*{ACKNOWLEDGEMENTS}

We would acknowledge the support by the Ministry of Science and
Technology of China under Grant Nos. 2016YFA0300404 and
2017YFA0403600, and the support by the National Natural Science
Foundation of China under Grants 11574285, 11504379, 11674327, and
U1532267. J.R.W. is also supported by the Natural Science Foundation
of Anhui Province under Grant 1608085MA19.

\appendix

\section{$\mathrm{Re}\Pi^{\mathrm{ret}}(\Omega,\mathbf{q})$ in the limit $\max(vq_{\bot},Aq_{z}^{2})\ll\Omega\ll\mu$
\label{App:PolaAppro}}

\subsection{$\mathrm{Re}\Pi_{++}^{\mathrm{ret}}(\Omega,\mathbf{q})$}

In the limit $\max(vq_{\bot},Aq_{z}^{2})\ll \Omega$, we have
\begin{eqnarray}
\mathrm{Re}\Pi_{++}^{\mathrm{ret}}(\Omega,\mathbf{q})
&\approx&-\frac{N}{16\pi^3}P\int d^3\mathbf{k}\nonumber
\\
&&\times\left(1+\frac{v^{2}k_{x}^{2}
+v^{2}k_{y}^{2}
+A^2k_{z}^{4}}{E_{\mathbf{k}}^{2}}\right)\nonumber
\\
&&\times\frac{\delta\left(\mu-E_{\mathbf{k}}\right)\frac{\partial
E_{\mathbf{k}}}{\partial k_{i}}q_{i}} {-\frac{\partial
E_{\mathbf{k}}}{\partial k_{j}}q_{j}+\Omega}\nonumber
\\
&=&-\frac{N}{8\pi^3}P\int d^3\mathbf{k}
\frac{\delta\left(\mu-E_{\mathbf{k}}\right)\frac{\partial
E_{\mathbf{k}}}{\partial k_{i}}\frac{q_{i}}{\Omega}}
{-\frac{\partial E_{\mathbf{k}}}{\partial
k_{j}}\frac{q_{j}}{\Omega}+1}\nonumber
\\
&=&-\frac{N}{8\pi^3}\int
d^3\mathbf{k}\delta\left(\mu-E_{\mathbf{k}}\right)\frac{\partial
E_{\mathbf{k}}}{\partial k_{i}}\frac{q_{i}}{\Omega}\nonumber
\\
&&\times\left(1+\frac{\partial E_{\mathbf{k}}}{\partial
k_{j}}\frac{q_{j}}{\Omega}\right).
\end{eqnarray}
Since $E_{\mathbf{k}} = \sqrt{v^{2}\left(k_{x}^{2}+v_{y}^{2}\right)
+ A^{2}k_{z}^{4}}$, it is clear that
\begin{eqnarray}
\frac{\partial E_{\mathbf{k}}}{\partial k_{i}}\frac{q_{i}}{\Omega}
&=& \frac{v^{2}k_{x}q_{x}+v^{2}k_{y}q_{y} +
2A^{2}k_{z}^{3}q_{z}}{E_{\mathbf{k}}\Omega}. \label{Eq:DiverEk}
\end{eqnarray}
Now $\mathrm{Re}\Pi_{++}^{\mathrm{ret}}(\Omega,\mathbf{q})$ can be
further written as
\begin{eqnarray}
\mathrm{Re}\Pi_{++}^{\mathrm{ret}}(\Omega,\mathbf{q}) &=& -
\frac{N}{8\pi^3}\int d^3\mathbf{k}\delta
\left(\mu-E_{\mathbf{k}}\right)\nonumber \\
&&\times\frac{v^{4}k_{x}^{2}q_{x}^{2} + v^{4}k_{y}^{2}q_{y}^{2} +
4A^{4}k_{z}^{6}q_{z}^{2}}{E_{\mathbf{k}}^{2}\Omega^{2}} \nonumber
\\
&=&-\frac{N}{8\pi^3}\int_{0}^{2\pi}d\varphi\int dk_{\bot}k_{\bot}dk_{z}
\delta\left(\mu-E_{\mathbf{k}}\right)\nonumber
\\
&&\times\frac{1}{E_{\mathbf{k}}^{2}\Omega^{2}}
\left[v^{4}k_{\bot}^{2}\cos^2\varphi q_{x}^{2}\right.\nonumber
\\
&&\left.+v^{4}k_{\bot}^{2}\sin^2\varphi q_{y}^{2} +
4A^{4}k_{z}^{6}q_{z}^{2}\right]\nonumber
\\
&=&-\frac{N}{4\pi^2}\left[\frac{v^{4}q_{\bot}^{2}}{\Omega^{2}}\int
dk_{\bot}d|k_{z}|k_{\bot}\delta\left(\mu-E_{\mathbf{k}}\right)
\right.\nonumber
\\
&&\times\frac{k_{\bot}^{2}}{E_{\mathbf{k}}^{2}} +
\frac{8A^{4}q_{z}^{2}}{\Omega^{2}}\int
dk_{\bot}d|k_{z}|k_{\bot}\nonumber
\\
&&\left.\times\delta\left(\mu-E_{\mathbf{k}}\right)
\frac{k_{z}^{6}}{E_{\mathbf{k}}^{2}}\right].
\end{eqnarray}
We then employ the transformations
\begin{eqnarray}
E = \sqrt{v^2k_{\bot}^{2}+A^{2}k_{z}^{4}},\qquad \delta =
\frac{Ak_{z}^{2}}{vk_{\bot}}, \label{Eq:TransformationA}
\end{eqnarray}
which are equivalent to
\begin{eqnarray}
k_{\bot}=\frac{E}{v\sqrt{1+\delta^{2}}},\qquad |k_{z}| =
\frac{\sqrt{\delta}\sqrt{E}}{\sqrt{A}\left(1 +
\delta^2\right)^{\frac{1}{4}}}. \label{Eq:TransformationB}
\end{eqnarray}
It is easy to verify that
\begin{eqnarray}
dk_{\bot}d|k_{z}|&=&
\left|\left|
\begin{array}{cc}
\frac{\partial k_{\bot}}{\partial E} & \frac{\partial
k_{\bot}}{\partial \delta}
\\
\frac{\partial q_{z}}{\partial E} & \frac{\partial k_{z}}{\partial
\delta}
\end{array}
\right|\right|dEd\delta\nonumber
\\
&=&\left|\frac{\partial k_{\bot}}{\partial E}\frac{\partial
k_{z}}{\partial \delta} - \frac{\partial k_{\bot}}{\partial
\delta}\frac{\partial k_{z}}{\partial E}\right|dEd\delta \nonumber
\\
&=&\frac{\sqrt{E}}{2v\sqrt{A}\sqrt{\delta}
\left(1+\delta^2\right)^{\frac{3}{4}}}dEd\delta.
\label{Eq:TransformationC}
\end{eqnarray}
Utilizing the transformations Eqs.~(\ref{Eq:TransformationB}) and
(\ref{Eq:TransformationC}), one can get
\begin{eqnarray}
\mathrm{Re}\Pi_{++}^{\mathrm{ret}}(\Omega,\mathbf{q}) &=&
-\frac{N}{4\pi^2}\left[\frac{q_{\bot}^{2}}{2\sqrt{A}\Omega^{2}}
\int_{0}^{+\infty} dE E^{\frac{3}{2}}
\delta\left(\mu-E\right)\right.\nonumber
\\
&&\times\int_{0}^{+\infty} d\delta\frac{1}{\sqrt{\delta}
\left(1+\delta^{2}\right)^{\frac{9}{4}}}\nonumber
\\
&&+\frac{4\sqrt{A}q_{z}^{2}}{v^{2}\Omega^{2}}\int_{0}^{+\infty}
dEE^{\frac{5}{2}}\delta\left(\mu-E\right)\nonumber
\\
&&\left.\times\int_{0}^{+\infty} d\delta
\frac{1}{\sqrt{\delta}\left(1 +
\delta^{2}\right)^{\frac{11}{4}}}\right]\nonumber
\\
&=&-\left(C_{++}^{\bot}\frac{q_{\bot}^{2}
\mu^{\frac{3}{2}}}{\Omega^{2}} + C_{++}^{z}\frac{q_{z}^{2}
\mu^{\frac{5}{2}}}{\Omega^{2}}\right),
\end{eqnarray}
where
\begin{eqnarray}
C_{++}^{\bot}&=&\frac{N}{5\pi^{2}\sqrt{A}},
\\
C_{++}^{z}&=&\frac{3N\Gamma\left(\frac{5}{4}\right)
\sqrt{A}}{2\pi^{\frac{3}{2}}\Gamma\left(\frac{11}{4}\right)v^{2}}.
\end{eqnarray}

\subsection{$\mathrm{Re}\Pi_{+-}^{\mathrm{ret}}(\Omega,\mathbf{q})$}

In the limit $\max(vq_{\bot},Aq_{z}^{2})\ll \Omega$,
$\mathrm{Re}\Pi_{+-}^{\mathrm{ret}}(\Omega,\mathbf{q})$ is
approximately given by
\begin{eqnarray}
\mathrm{Re}\Pi_{+-}^{\mathrm{ret}}(\Omega,\mathbf{q}) &\approx&
\frac{N}{16\pi^3}\int d^3\mathbf{k} \left[1 -
\frac{F_{\mathbf{k},\mathbf{q}}}{E_{\mathbf{k}}\left(\frac{\partial
E_{\mathbf{k}}}{\partial k_{i}}q_{i}+E_{\mathbf{k}}\right)}\right]
\nonumber \\
&&\times\frac{\theta\left(E_{\mathbf{k}} -
\mu\right)}{2E_{\mathbf{k}}+\Omega}\nonumber
\\
&\approx&\frac{N}{16\pi^3}\int d^3\mathbf{k} \left[1 -
\frac{F_{\mathbf{k},\mathbf{q}}}{E_{\mathbf{k}}^{2}}\right.\nonumber
\\
&&\left.\left(1-\frac{\partial E_{\mathbf{k}}}{\partial k_{i}}
\frac{q_{i}}{E_{\mathbf{k}}}\right)\right]
\frac{\theta\left(E_{\mathbf{k}}-\mu\right)}
{2E_{\mathbf{k}}+\Omega}. \label{Eq:PolaPNAppAxppA}
\end{eqnarray}
Substituting Eq.~(\ref{Eq:DiverEk}) into
Eq.~(\ref{Eq:PolaPNAppAxppA}), one gets
\begin{eqnarray}
\mathrm{Re}\Pi_{+-}^{\mathrm{ret}}(\Omega,\mathbf{q}) &=&
\frac{N}{16\pi^3}\int d^3\mathbf{k} \left[-\frac{A^2
k_{z}^{2}q_{z}^{2}}{E_{\mathbf{k}}^{2}}\right.\nonumber
\\
&&\left.+\frac{v^{4}k_{x}^{2}q_{x}^{2}+v^{2}k_{y}^{2}q_{y}^{2} +
4A^4k_{z}^{6}q_{z}^{2}}{E_{\mathbf{k}}^{4}} \right]\nonumber
\\
&&\times\frac{\theta\left(E_{\mathbf{k}} -
\mu\right)}{2E_{\mathbf{k}}+\Omega}.\label{Eq:PolaPNAppAxppB}
\end{eqnarray}
Performing the integration of azimuth angle, we obtain
\begin{eqnarray}
\mathrm{Re}\Pi_{+-}^{\mathrm{ret}}(\Omega,\mathbf{q}) &=&
\frac{N}{8\pi^2}\left[v^{4}q_{\bot}^{2}\int dk_{\bot}
d|k_{z}|k_{\bot}\frac{k_{\bot}^{2}}{E_{\mathbf{k}}^{4}}\right.\nonumber
\\
&&\frac{\theta\left(E_{\mathbf{k}}-\mu\right)}{2E_{\mathbf{k}} +
\Omega}+2A^{2}q_{z}^{2}\int dk_{\bot}d|k_{z}|k_{\bot}\nonumber
\\
&&\times\left(-\frac{ k_{z}^{2}}{E_{\mathbf{k}}^{2}} + \frac{ 4A^2
k_{z}^{6}}{E_{\mathbf{k}}^{4}}\right)\nonumber \\
&&\left.\times\frac{\theta\left(E_{\mathbf{k}}-\mu\right)}
{2E_{\mathbf{k}}+\Omega}\right].  \label{Eq:PolaPNAppAxppC}
\end{eqnarray}
Using the transformations Eqs.~(\ref{Eq:TransformationB}) and
(\ref{Eq:TransformationC}), and performing the integration of
$\delta$, $\mathrm{Re}\Pi_{+-}^{\mathrm{ret}}(\Omega,\mathbf{q})$
can be further written as
\begin{eqnarray}
\mathrm{Re}\Pi_{+-}^{\mathrm{ret}}(\Omega,\mathbf{q}) &=&
\frac{N}{8\pi^2}\left[\frac{4q_{\bot}^{2}}{5\sqrt{A}}\int_{\mu}^{\Lambda}
dE \frac{1}{E^{\frac{1}{2}}}\frac{1} {2E+\Omega}\right.\nonumber
\\
&&+\frac{\sqrt{A}q_{z}^{2}}{v^{2}}\int_{\mu}^{\Lambda} dE
E^{\frac{1}{2}}\frac{1} {2E+\Omega} \nonumber
\\
&&\left.\times\left(\frac{6}{7}-\frac{3\sqrt{\pi}
\Gamma\left(\frac{5}{4}\right)}{2\Gamma\left(\frac{11}{4}\right)}\right)\right].
\label{Eq:PolaPNAppAxppD}
\end{eqnarray}
In the regime $\Omega \ll \mu$, it can be approximated by
\begin{eqnarray}
\mathrm{Re}\Pi_{+-}^{\mathrm{ret}}(\Omega,\mathbf{q}) &\approx&
C_{+-}^{\bot}q_{\bot}^{2}+C_{+-}^{z}q_{z}^{2},
\label{Eq:PolaPNAppAxppE}
\end{eqnarray}
where
\begin{eqnarray}
C_{+-}^{\bot} &=& \frac{N}{20\pi^2
\sqrt{A}}\left(\frac{1}{\Lambda^{\frac{1}{2}}} -
\frac{1}{\mu^{\frac{1}{2}}}\right), \label{Eq:CPNBotApp}
\\
C_{+-}^{z}&=&\left(\frac{6}{7}-\frac{3\sqrt{\pi}
\Gamma\left(\frac{5}{4}\right)}{2\Gamma\left(\frac{11}{4}\right)}\right)
\frac{N\sqrt{A}}{32v^{2}}\nonumber
\\
&&\times\left(\Lambda^{\frac{1}{2}}-\mu^{\frac{1}{2}}\right).
\label{Eq:CPNZApp}
\end{eqnarray}

\subsection{$\mathrm{Re}\Pi_{-+}^{\mathrm{ret}}(\Omega,\mathbf{q})$}

In the limit $\max(vq_{\bot},Aq_{z}^{2})\ll \Omega$,
$\mathrm{Re}\Pi_{-+}^{\mathrm{ret}}(\Omega,\mathbf{q})$ takes the
form
\begin{eqnarray}
\mathrm{Re}\Pi_{-+}^{\mathrm{ret}}(\Omega,\mathbf{q}) &\approx&
\frac{N}{16\pi^3}\int d^3\mathbf{k} \left[1 -
\frac{F_{\mathbf{k},\mathbf{q}}}{E_{\mathbf{k}} \left(\frac{\partial
E_{\mathbf{k}}}{\partial k_{i}}q_{i} +
E_{\mathbf{k}}\right)}\right]\nonumber
\\
&&\times\frac{\theta\left(E_{\mathbf{k}} -
\mu\right)}{2E_{\mathbf{k}}-\Omega}\nonumber
\\
&\approx&\frac{N}{16\pi^3}\int d^3\mathbf{k}
\left[1-\frac{F_{\mathbf{k},\mathbf{q}}}{E_{\mathbf{k}}^{2}}\right.\nonumber
\\
&&\left.\left(1-\frac{\partial E_{\mathbf{k}}}{\partial k_{i}}
\frac{q_{i}}{E_{\mathbf{k}}}\right)\right]\frac{\theta
\left(E_{\mathbf{k}}-\mu\right)}
{2E_{\mathbf{k}}-\Omega},\label{Eq:PolaNPAppAxppA}
\end{eqnarray}
Substituting Eq.~(\ref{Eq:DiverEk}) into
Eq.~(\ref{Eq:PolaNPAppAxppA}), and carrying out the integration of
azimuth angle, we can get
\begin{eqnarray}
\mathrm{Re}\Pi_{-+}^{\mathrm{ret}}(\Omega,\mathbf{q}) &=&
\frac{N}{8\pi^2}\left[v^{4}q_{\bot}^{2}\int dk_{\bot}
d|k_{z}|k_{\bot}\frac{k_{\bot}^{2}}{E_{\mathbf{k}}^{4}}
\frac{\theta\left(E_{\mathbf{k}}-\mu\right)}{2E_{\mathbf{k}} -
\Omega}\right.\nonumber \\
&&+2A^{2}q_{z}^{2}\int dk_{\bot}d|k_{z}|k_{\bot}\nonumber
\\
&&\left.\times\left(-\frac{ k_{z}^{2}}{E_{\mathbf{k}}^{2}} + \frac{
4A^2 k_{z}^{6}}{E_{\mathbf{k}}^{4}}\right)
\frac{\theta\left(E_{\mathbf{k}}-\mu\right)}{2E_{\mathbf{k}} -
\Omega}\right]. \label{Eq:PolaNPAppAxppB}
\end{eqnarray}
Utilizing the transformations Eqs.~(\ref{Eq:TransformationB}) and
(\ref{Eq:TransformationC}), and performing the integration of
$\delta$, we can get
\begin{eqnarray}
\mathrm{Re}\Pi_{-+}^{\mathrm{ret}}(\Omega,\mathbf{q}) &=&
\frac{N}{8\pi^2}\left[\frac{4q_{\bot}^{2}}{5\sqrt{A}}\int_{\mu}^{\Lambda}
dE \frac{1}{E^{\frac{1}{2}}}\frac{1} {2E-\Omega}\right.\nonumber
\\
&&+\frac{\sqrt{A}q_{z}^{2}}{v^{2}}\int_{\mu}^{\Lambda} dE
E^{\frac{1}{2}}\frac{1} {2E-\Omega} \nonumber
\\
&&\left.\times\left(\frac{6}{7}-\frac{3\sqrt{\pi}
\Gamma\left(\frac{5}{4}\right)}{2\Gamma\left(\frac{11}{4}\right)}\right)\right].
\label{Eq:PolaNPAppAxppC}
\end{eqnarray}
In the regime $\Omega \ll \mu$, it can be approximately written as
\begin{eqnarray}
\mathrm{Re}\Pi_{-+}^{\mathrm{ret}}(\Omega,\mathbf{q}) &\approx&
C_{-+}^{\bot}q_{\bot}^{2} + C_{-+}^{z}q_{z}^{2},
\label{Eq:PolaNPAppAxppD}
\end{eqnarray}
where
\begin{eqnarray}
C_{-+}^{\bot}&=&\frac{N}{20\pi^2\sqrt{A}}
\left(\frac{1}{\Lambda^{\frac{1}{2}}} -
\frac{1}{\mu^{\frac{1}{2}}}\right), \label{Eq:CNPBotApp}
\\
C_{-+}^{z}&=&\left(\frac{6}{7}-\frac{3\sqrt{\pi}
\Gamma\left(\frac{5}{4}\right)}{2\Gamma\left(\frac{11}{4}\right)}\right)
\frac{N\sqrt{A}}{32v^{2}}\nonumber \\
&&\times\left(\Lambda^{\frac{1}{2}} - \mu^{\frac{1}{2}}\right).
\label{Eq:CNPZApp}
\end{eqnarray}
Comparing Eqs.~(\ref{Eq:PolaPNAppAxppE})-(\ref{Eq:CPNZApp}) with
Eqs.~(\ref{Eq:PolaNPAppAxppD})-(\ref{Eq:CNPZApp}), we know that
\begin{eqnarray}
\mathrm{Re}\Pi_{+-}^{\mathrm{ret}}(\Omega,\mathbf{q}) \approx
\mathrm{Re}\Pi_{-+}^{R}(\Omega,\mathbf{q})
\end{eqnarray}
in the regime $\max(vq_{\bot},Aq_{z}^{2})\ll \Omega \ll \mu$.

\section{The density of fermion \label{App:DensityFermion}}

The density of fermion is given by
\begin{eqnarray}
n_{e} &=& \int\frac{d^3\mathbf{k}}{(2\pi)^{3}}
\theta\left(\mu-E_{\mathbf{k}}\right)\nonumber
\\
&=&\frac{1}{8\pi^{3}}\int_{0}^{2\pi}d\varphi\int dk_{\bot}k_{\bot}
dk_{z}\theta\left(\mu-E_{\mathbf{k}}\right)\nonumber
\\
&=&\frac{1}{2\pi^{2}}\int dk_{\bot}k_{\bot}
d|k_{z}|\theta\left(\mu-E_{\mathbf{k}}\right).
\end{eqnarray}
where $E_{\mathbf{k}} = \sqrt{v^{2}k_{\bot}^{2}+A^{2}k_{z}^{4}}$.
Employing the transformations Eqs.~(\ref{Eq:TransformationB}) and
(\ref{Eq:TransformationC}), the density becomes
\begin{eqnarray}
n_{e} &=& \frac{1}{4\pi^{2}v^{2}\sqrt{A}}\int_{0}^{+\infty}dE
\int_{0}^{+\infty}d\delta \frac{E^{\frac{3}{2}}}{\sqrt{\delta}
\left(1+\delta^2\right)^{\frac{5}{4}}}
\theta\left(\mu-E\right)\nonumber
\\
&=&\frac{1}{4\pi^{2}v^{2}\sqrt{A}}\int_{0}^{\mu}dE
E^{\frac{3}{2}}\int_{0}^{+\infty}d\delta \frac{1}{\sqrt{\delta}
\left(1+\delta^2\right)^{\frac{5}{4}}}\nonumber
\\
&=&\frac{1}{5\pi^{2}v^{2}\sqrt{A}}\mu^{\frac{5}{2}}.
\end{eqnarray}

\begin{widetext}

\section{Further Calculation of polarization function \label{App:PolaAnagleIntegral}}

\subsection{$\mathrm{Im}\Pi^{\mathrm{ret}}(\Omega,\mathbf{q})$}

Eq.~(\ref{Eq:ImPolaMainText}) in the main body of the paper can be
further written as
\begin{eqnarray}
\mathrm{Im}\Pi^{\mathrm{ret}}(\Omega,\mathbf{q}) =
I_{1}-I_{2}+I_{3}+I_{4}-I_{5}-I_{6}+I_{7}+I_{8}-I_{9}-I_{10}.
\end{eqnarray}
It can be found that
\begin{eqnarray}
I_{2}&=&I_{1}(\Omega\rightarrow-\Omega). \label{Eq:RelationIA}\\
I_4 &=& I_3\left(\mu\rightarrow-\mu\right),\qquad I_5 =
I_3\left(\Omega\rightarrow-\Omega\right),\qquad I_6 =
I_3\left(\Omega\rightarrow-\Omega,\mu\rightarrow-\mu\right),
\label{Eq:RelationIB} \\
I_8 &=& I_7\left(\mu\rightarrow-\mu\right),\qquad I_9 =
I_7\left(\Omega\rightarrow-\Omega\right),\qquad I_{10} =
I_7\left(\Omega\rightarrow-\Omega,\mu\rightarrow-\mu\right).
\label{Eq:RelationIC}
\end{eqnarray}
Therefore, we only need to calculate $I_1$, $I_3$, and $I_7$, which
are given by
\begin{eqnarray}
I_{1}&=&-\frac{N}{16\pi^{2}}\int dk_{\bot}k_{\bot}\int
dk_{z}\int_{0}^{2\pi} d\varphi\left[1 -
\frac{F_{\mathbf{k},\mathbf{q}}}{E_{\mathbf{k}}E_{\mathbf{k} +
\mathbf{q}}}\right]\delta\left(E_{\mathbf{k}}+E_{\mathbf{k} +
\mathbf{q}}+\Omega\right), \label{Eq:ExpressionI1}
\\
I_{3}&=&\frac{N}{16\pi^{2}}\int dk_{\bot}k_{\bot}\int dk_{z}
n_F\left(E_{\mathbf{k}}-\mu\right)\int_{0}^{2\pi} d\varphi
\left[1+\frac{F_{\mathbf{k},\mathbf{q}}}{E_{\mathbf{k}}E_{\mathbf{k}
+ \mathbf{q}}}\right] \delta\left(E_{\mathbf{k}} -
E_{\mathbf{k}+\mathbf{q}}+\Omega\right), \label{Eq:ExpressionI2}
\\
I_{7}&=&\frac{N}{16\pi^{2}}\int dk_{\bot}k_{\bot}\int dk_{z}
n_F\left(E_{\mathbf{k}} - \mu\right)\int_{0}^{2\pi} d\varphi \left[1
- \frac{F_{\mathbf{k},\mathbf{q}}}{E_{\mathbf{k}}E_{\mathbf{k} +
\mathbf{q}}}\right]\delta\left(E_{\mathbf{k}} +
E_{\mathbf{k}+\mathbf{q}}+\Omega\right), \label{Eq:ExpressionI3}
\end{eqnarray}
where
\begin{eqnarray}
F_{\mathbf{k},\mathbf{q}} &=& v^{2}k_{\bot}^{2} +
v^{2}k_{\bot}q_{\bot}\cos(\varphi) + A^2k_{z}^{2}
\left(k_{z}+q_{z}\right)^{2},
\\
E_{\mathbf{k}+\mathbf{q}} &=& \sqrt{v^{2}\left(k_{\bot}^{2} +
q_{\bot}^{2}+2k_{\bot}q_{\bot}\cos(\varphi)\right) +
A^{2}\left(k_{z}+q_{z}\right)^{4}}.
\end{eqnarray}
Suppose that
\begin{eqnarray}
\delta\left[F_{1}(\varphi)\right] = \delta\left(E_{\mathbf{k}} +
E_{\mathbf{k}+\mathbf{q}}+\Omega\right) =
\delta\left(\sqrt{v^{2}k_{\bot}^{2} +
A^{2}k_{z}^{4}}+\sqrt{v^{2}\left(k_{\bot}^{2} +
q_{\bot}^{2}+2k_{\bot}q_{\bot}\cos(\varphi)\right) +
A^{2}\left(k_{z}+q_{z}\right)^{4}}+\Omega\right).
\end{eqnarray}
After tedious evaluation, we get
\begin{eqnarray}
\delta\left[F_{1}(\varphi)\right] &=& 2\left(E_{\mathbf{k}} +
\Omega\right)\left[\left(E_{\mathbf{k}}+\Omega\right)^{2} -
\left(E_{1}\left(\mathbf{k},\mathbf{q}\right)\right)^{2}\right]^{-\frac{1}{2}}
\left[-\left(E_{\mathbf{k}}+\Omega\right)^{2} +
\left(E_{2}(\mathbf{k},\mathbf{q})\right)^{2}\right]^{-\frac{1}{2}}
\left[\delta\left(\varphi-\varphi_{1}^{a}\right)+\delta(\varphi -
\varphi_{2}^{a})\right] \nonumber
\\
&&\times\theta\left(-E_{\mathbf{k}}-\Omega\right)
\theta\left(-E_{\mathbf{k}} - \Omega -
E_{1}(\mathbf{k},\mathbf{q})\right)
\theta\left(E_{2}(\mathbf{k},\mathbf{q}) +
E_{\mathbf{k}}+\Omega\right), \label{Eq:ExpressionDeltaF1}
\end{eqnarray}
where
\begin{eqnarray}
E_{1}(\mathbf{k},\mathbf{q}) &=& \sqrt{v^{2}\left(k_{\bot} -
q_{\bot}\right)^{2}+A^{2}\left(k_{z} + q_{z}\right)^{4}},
\\
E_{2}(\mathbf{k},\mathbf{q})&=&\sqrt{v^{2}\left(k_{\bot} +
q_{\bot}\right)^{2}+A^{2}\left(k_{z}+q_{z}\right)^{4}}.
\end{eqnarray}
Here, $\varphi_{1}^{a}$ and $\varphi_{2}^{b}$ are the two angles
satisfying $F_{1}(\varphi_{1,2}^{a})=0$ in the range $(0,2\pi)$. We
then let
\begin{eqnarray}
\delta\left[F_{2}(\varphi)\right] =
\delta\left(E_{\mathbf{k}}-E_{\mathbf{k}+\mathbf{q}}+\Omega\right) =
\delta\left(\sqrt{v^{2}k_{\bot}^{2}+A^{2}k_{z}^{4}} - \sqrt{v^{2}
\left(k_{\bot}^{2}+q_{\bot}^{2}+2k_{\bot}q_{\bot}\cos(\varphi)\right)
+ A^{2}\left(k_{z}+q_{z}\right)^{4}}+\Omega\right).
\end{eqnarray}
We similarly obtain
\begin{eqnarray}
\delta\left(F_{2}(\varphi)\right) &=& 2\left(E_{\mathbf{k}} +
\Omega\right)\left[\left(E_{\mathbf{k}}+\Omega\right)^{2} -
\left(E_{1}(\mathbf{k},\mathbf{q})\right)^{2}\right]^{-\frac{1}{2}}
\left[-\left(E_{\mathbf{k}}+\Omega\right)^{2} +
\left(E_{2}(\mathbf{k},\mathbf{q})\right)^{2}\right]^{-\frac{1}{2}}
\left[\delta\left(\varphi-\varphi_{1}^{b}\right) +
\delta(\varphi-\varphi_{2}^{b})\right]\nonumber
\\
&&\times\theta\left(E_{\mathbf{k}}+\Omega\right)
\theta\left(E_{\mathbf{k}}+\Omega -
E_{1}(\mathbf{k},\mathbf{q})\right)
\theta\left(E_{2}(\mathbf{k},\mathbf{q})-\Omega\right),
\label{Eq:ExpressionDeltaF2}
\end{eqnarray}
where $\varphi_{1}^{b}$ and $\varphi_{2}^{b}$ satisfy
$F_{2}(\varphi_{1,2}^{b})=0$ in the range $(0,2\pi)$. After
substituting Eq.~(\ref{Eq:ExpressionDeltaF1}) into
Eqs.~(\ref{Eq:ExpressionI1}) and (\ref{Eq:ExpressionI3}), and also
substituting Eq.~(\ref{Eq:ExpressionDeltaF2}) into
Eq.~(\ref{Eq:ExpressionI2}), and then carrying out the integration
over $\varphi$, we can get the simplified expressions of $I_{1}$,
$I_{2}$, and $I_{3}$. Making use of the relations
Eqs.~(\ref{Eq:RelationIA}-(\ref{Eq:RelationIC}), we eventually have
\begin{eqnarray}
\mathrm{Im}\Pi^{\mathrm{ret}}(\Omega,\mathbf{q}) &=&
\sum_{\alpha=\pm1}\mathrm{sgn}(\Omega)\frac{N}{8\pi^{2}}\int
dk_{\bot}k_{\bot}\int dk_{z}\left[\delta_{1,\alpha} -
n_F\left(E(\mathbf{k})+\alpha\mu\right)\right]
\frac{1}{E(\mathbf{k})}F_{A1}(\mathbf{k},\mathbf{q})
F_{B1}(\mathbf{k},\mathbf{q})H_{1}(\mathbf{k},\mathbf{q})\nonumber
\\
&&+\sum_{\alpha=\pm1}\mathrm{sgn}(\Omega)\frac{N}{8\pi^{2}}\int
dk_{\bot}k_{\bot}\int dk_{z} n_F\left(E(\mathbf{k}) +
\alpha\mu\right)\frac{1}{E(\mathbf{k})}\nonumber
\\
&&\times\left\{F_{A2}(\mathbf{k},\mathbf{q})F_{B2}
\left(\mathbf{k},\mathbf{q}\right)H_{2}(\mathbf{k},\mathbf{q}) -
F_{A3}(\mathbf{k},\mathbf{q}) F_{B1}(\mathbf{k},\mathbf{q})
H_{3}(\mathbf{k},\mathbf{q})\right\},
\end{eqnarray}
where
\begin{eqnarray}
F_{A1}\left(\mathbf{k},\mathbf{q}\right) &=&
\left[\left(-E(\mathbf{k})+\left|\Omega\right|\right)^{2} -
\left(E_{1}(\mathbf{k},\mathbf{q})\right)^{2}\right]^{-\frac{1}{2}}
\left[-\left(-E(\mathbf{k})+\left|\Omega\right|\right)^{2} +
\left(E_{2}(\mathbf{k},\mathbf{q})\right)^{2}\right]^{-\frac{1}{2}},
\\
F_{A2}(\mathbf{k},\mathbf{q})&=&\left[\left(E(\mathbf{k}) +
\left|\Omega\right|\right)^{2} -
\left(E_{1}(\mathbf{k},\mathbf{q})\right)^{2}\right]^{-\frac{1}{2}}
\left[-\left(E(\mathbf{k})+\left|\Omega\right|\right)^{2} +
\left(E_{2}(\mathbf{k},\mathbf{q})\right)^{2}\right]^{-\frac{1}{2}},
\\
F_{A3}\left(\mathbf{k},\mathbf{q}\right)&=&\left[\left(E(\mathbf{k})
- \left|\Omega\right|\right)^{2} -
\left(E_{1}(\mathbf{k},\mathbf{q})\right)^{2}\right]^{-\frac{1}{2}}
\left[-\left(E(\mathbf{k})-\left|\Omega\right|\right)^{2} +
\left(E_{2}(\mathbf{k},\mathbf{q})\right)^{2}\right]^{-\frac{1}{2}},
\\
F_{B1}\left(\mathbf{k},\mathbf{q}\right)&=&\left[
\left(2E(\mathbf{k})-\left|\Omega\right|\right)^{2}
-v^{2}q_{\bot}^{2}-A^{2}\left(2k_{z}+q_{z}\right)^{2}q_{z}^{2}
\right],
\\
F_{B2}\left(\mathbf{k},\mathbf{q}\right)&=&\left[
\left(2E(\mathbf{k})+\left|\Omega\right|\right)^{2}
-v^{2}q_{\bot}^{2}-A^{2}\left(2k_{z}+q_{z}\right)^{2}q_{z}^{2}
\right],
\\
H_{1}(\mathbf{k},\mathbf{q})&=&\theta\left(-E(\mathbf{k}) +
\left|\Omega\right|\right)\theta\left(-E(\mathbf{k})+\left|\Omega\right|
- E_{1}(\mathbf{k},\mathbf{q})\right)
\theta\left(E_{2}(\mathbf{k},\mathbf{q}) + E(\mathbf{k}) -
\left|\Omega\right|\right),
\\
H_{2}(\mathbf{k},\mathbf{q})&=&\theta\left(E(\mathbf{k}) +
\left|\Omega\right|\right)\theta\left(E(\mathbf{k})+\left|\Omega\right|
-E_{1}(\mathbf{k},\mathbf{q})\right)\theta\left(E_{2}(\mathbf{k},\mathbf{q})
- E(\mathbf{k})-\left|\Omega\right|\right),
\\
H_{3}(\mathbf{k},\mathbf{q})&=&\theta\left(E(\mathbf{k}) -
\left|\Omega\right|\right)\theta\left(E(\mathbf{k})-\left|\Omega\right|
- E_{1}(\mathbf{k},\mathbf{q})\right)
\theta\left(E_{2}(\mathbf{k},\mathbf{q})-E(\mathbf{k}) +
\left|\Omega\right|\right).
\end{eqnarray}

\subsection{$\mathrm{Re}\Pi^{\mathrm{ret}}(\Omega,\mathbf{q})$}

After further calculations, we write Eq.~(\ref{Eq:RePolaMainText})
of the main body of the paper in the form
\begin{eqnarray}
\mathrm{Re}\Pi^{\mathrm{ret}}(\Omega,\mathbf{q}) =
J_{1}+J_{2}+J_{3}+J_{4}+J_{5}+J_{6}.
\end{eqnarray}
There exists a number of identities:
\begin{eqnarray}
J_2 &=& J_1\left(\Omega\rightarrow - \Omega\right),
\label{Eq:RelationJA} \\
J_4 &=& J_3\left(\mu\rightarrow-\mu\right),\qquad J_5 =
J_3\left(\Omega\rightarrow-\Omega\right),\qquad J_6 =
J_3\left(\mu\rightarrow-\mu,\Omega\rightarrow-\Omega\right).
\label{Eq:RelationJB}
\end{eqnarray}
It is thus only necessary to calculate $J_1$ and $J_3$, given by
\begin{eqnarray}
J_{1} &=& \frac{N}{8\pi^{3}}\int dk_{\bot}k_{\bot}\int
dk_{z}\frac{1}{E_{\mathbf{k}}}M_{1},\label{Eq:ExpressionJ1}
\\
J_{3} &=& -\frac{N}{8\pi^{3}}\int dk_{\bot}k_{\bot}\int dk_{z}
n_F\left(E_{\mathbf{k}}-\mu\right)\frac{1}{E_{\mathbf{k}}}M_{1},
\label{Eq:ExpressionJ2}
\end{eqnarray}
where
\begin{eqnarray}
M_{1} = \mathcal{P}\int_{0}^{2\pi}d\varphi \frac{E_{\mathbf{k}}
\left(E_{\mathbf{k}}+\Omega\right)+v^{2}k_{\bot}^{2} + v^{2}
k_{\bot}q_{\bot}\cos(\varphi) + A^2k_{z}^{2} \left(k_{z} +
q_{z}\right)^{2}}{\left(E_{\mathbf{k}}+\Omega\right)^2 -
\left(v^{2}\left(k_{\bot}^{2}+q_{\bot}^{2}+2k_{\bot}q_{\bot}
\cos(\varphi)\right)+A^{2}\left(k_{z}+q_{z}\right)^{4}\right)}.
\end{eqnarray}
It is convenient to define $Z=e^{i\varphi}$, which implies that
\begin{eqnarray}
\cos\varphi = \frac{e^{i\varphi} + e^{-i\varphi}}{2} = \frac{Z +
Z^{-1}}{2},\qquad d\varphi = \frac{dZ}{iZ}.
\end{eqnarray}
Now $M_{1}$ can be expressed as
\begin{eqnarray}
M_{1} = \frac{1}{i}\mathcal{P}\oint_{|Z| = 1}dZ F(Z),
\end{eqnarray}
where
\begin{eqnarray}
F(Z) = \frac{E_{\mathbf{k}}\left(E_{\mathbf{k}} +
\Omega\right)Z+v^{2}k_{\bot}^{2}Z + \frac{1}{2}
v^{2}k_{\bot}q_{\bot}\left(Z^{2}+1\right) + A^2k_{z}^{2}
\left(k_{z}+q_{z}\right)^{2}Z}{Z \left[\left(E_{\mathbf{k}} +
\Omega\right)^2 Z - \left(v^{2}\left(k_{\bot}^{2} +
q_{\bot}^{2}\right)Z + v^{2}k_{\bot}q_{\bot} \left(Z^{2} + 1\right)
+ A^{2}\left(k_{z}+q_{z}\right)^{4}Z\right)\right]}.
\end{eqnarray}
Using the residue theorem, we obtain
\begin{eqnarray}
M_{1} &=& -\pi+\pi \left[-\left(3E_{\mathbf{k}} + \Omega\right)
\left(E_{\mathbf{k}}+\Omega\right) - v^{2}\left(k_{\bot}^{2} -
q_{\bot}^{2}\right) - A^{2}\left(-2k_{z}^{2} + \left(k_{z} +
q_{z}\right)^{2}\right)\left(k_{z} +
q_{z}\right)^{2}\right]\nonumber
\\
&&\times\left[\left(\left(E_{2}(\mathbf{k},\mathbf{q})\right)^{2} -
\left(E_{\mathbf{k}}+\Omega\right)^2\right)
\left(\left(E_{1}\left(\mathbf{k},\mathbf{q}\right)\right) -
\left(E_{\mathbf{k}}+\Omega\right)^2\right)\right]^{-\frac{1}{2}}\nonumber
\\
&&\times\left[\theta\left(\left(E_{2}(\mathbf{k},\mathbf{q})\right)^{2}
- \left(E_{\mathbf{k}}+\Omega\right)^2\right)
\theta\left(\left(E_{1}\left(\mathbf{k},\mathbf{q}\right)\right) -
\left(E_{\mathbf{k}}+\Omega\right)^2\right)\right.\nonumber
\\
&&\left.-\theta\left(\left(E_{\mathbf{k}}+\Omega\right)^2 -
\left(E_{2}(\mathbf{k},\mathbf{q})\right)^{2}\right)
\theta\left(\left(E_{\mathbf{k}}+\Omega\right)^2 -
\left(E_{1}(\mathbf{k},\mathbf{q})\right)^{2}\right)\right].
\label{Eq:ExpressionM1}
\end{eqnarray}
Substituting Eq.~(\ref{Eq:ExpressionM1}) into
Eqs.~(\ref{Eq:ExpressionJ1}) and (\ref{Eq:ExpressionJ2}), it is
straightforward to get the expressions of $J_{1}$ and $J_{3}$. With
help of (\ref{Eq:RelationJA}) and (\ref{Eq:RelationJB}), we finally
find that
\begin{eqnarray}
\mathrm{Re}\Pi^{\mathrm{ret}}(\Omega,\mathbf{q}) &=&
\frac{N}{4\pi^{2}}\sum_{\alpha=\pm1}\int dk_{\bot}k_{\bot}\int
dk_{z} n_F\left(E_{\mathbf{k}}+\alpha\mu\right)
\frac{1}{E_{\mathbf{k}}} \nonumber \\
&&+\frac{N}{8\pi^{2}}\sum_{\alpha = \pm 1} \sum_{\alpha'=\pm1}\int
dk_{\bot}k_{\bot}\int dk_{z} \left[\delta_{1\alpha} -
n_F\left(E_{\mathbf{k}}+\alpha\mu\right)\right]
\frac{1}{E_{\mathbf{k}}}\nonumber
\\
&&\times\left[-\left(3E_{\mathbf{k}}+\alpha\Omega'\right)
\left(E_{\mathbf{k}}+\alpha'\Omega\right)
-v^{2}\left(k_{\bot}^{2}-q_{\bot}^{2}\right) +
A^{2}\left(-2k_{z}^{2}+\left(k_{z}+q_{z}\right)^{2}\right)
\left(k_{z}+q_{z}\right)^{2}\right]\nonumber
\\
&&\times\left[\left(\left(E_{1}(\mathbf{k},\mathbf{q})\right)^{2} -
\left(E_{\mathbf{k}}+\alpha'\Omega\right)^2\right)
\left(\left(E_{2}(\mathbf{k},\mathbf{q})\right)^{2} -
\left(E_{k}+\alpha'\Omega\right)^2\right)\right]^{-\frac{1}{2}}\nonumber
\\
&&\times\left[\theta\left(\left(E_{1}(\mathbf{k},\mathbf{q})\right)^{2}
- \left(E_{k}+\alpha'\Omega\right)^2\right)
\theta\left(\left(E_{2}(\mathbf{k},\mathbf{q})\right)^{2} -
\left(E_{k}+\alpha'\Omega\right)^2\right) \right.\nonumber
\\
&&\left.-\theta\left(\left(E_{\mathbf{k}}+\alpha'\Omega\right)^2 -
\left(E_{1}(\mathbf{k},\mathbf{q})\right)^{2}\right)
\theta\left(\left(E_{\mathbf{k}}+\alpha'\Omega\right)^2 -
\left(E_{2}(\mathbf{k},\mathbf{q})\right)^{2}\right)\right].
\end{eqnarray}
In the derivation, a constant term that is independent of $\Omega$,
$\mathbf{q}$, $\mu$, and $T$ has been dropped.

\section{Debye Screening\label{App:Screening}}

At $\Omega = 0$ and $|\mathbf{q}| = 0$, we have
\begin{eqnarray}
\mathrm{Re}\Pi^{\mathrm{ret}}(0,0) = \frac{N}{2\pi^{2}} \sum_{\alpha
= \pm 1}\int dk_{\bot}d|k_{z}|k_{\bot} n_F\left(E_{\mathbf{k}} +
\alpha\mu\right)\frac{1}{E_{\mathbf{k}}}.
\end{eqnarray}
Adopting the transformations shown in
Eqs.~(\ref{Eq:TransformationB}) and (\ref{Eq:TransformationC}) gives
rise to
\begin{eqnarray}
\mathrm{Re}\Pi^{\mathrm{ret}}(0,0) &=&
\frac{N}{2\pi^{2}v^{2}\sqrt{A}}\sum_{\alpha = \pm
1}\int_{0}^{+\infty} dE \sqrt{E} \frac{1}{e^{\frac{E +
\alpha\mu}{T}}+1}\nonumber
\\
&=& \frac{NT^{\frac{3}{2}}}{2\pi^{2}v^{2}\sqrt{A}} \sum_{\alpha =
\pm 1}\int_{0}^{+\infty} dx\sqrt{x} \frac{1}{e^{x +
\frac{\alpha\mu}{T}}+1}.
\end{eqnarray}

\end{widetext}

\end{document}